\documentclass[preprint]{elsarticle}
\usepackage{epsfig}
\usepackage{graphicx}
\usepackage{amssymb}
\newcommand{\AmS}{{\protect\the\textfont2
  A\kern-.1667em\lower.5ex\hbox{M}\kern-.125emS}}
\newcommand{\ba}{\begin{array}}
\newcommand{\ea}{\end{array}}

\def\beq{\begin{equation}}   
\def\eeq{\end{equation}}

\def\bea{\begin{eqnarray}}
\def\eea{\end{eqnarray}}

\def\beq{\begin{equation}}   
\def\eeq{\end{equation}}

\def\bea{\begin{eqnarray}}
\def\eea{\end{eqnarray}}

\begin{document}


\title{\Large Searching for hidden sector in multiparticle production at 
LHC}




\author[ific]{Miguel-Angel Sanchis-Lozano}
\ead{Miguel.Angel.Sanchis@ific.uv.es}

\author[cern,uta]{Edward K. Sarkisyan-Grinbaum\corref{cor1}}
\ead{sedward@cern.ch}

\author[ddi]{Salvador Moreno-Picot}\ead{Salvador.Moreno@uv.es}

\address[ific]{Instituto de F\'{\i}sica Corpuscular (IFIC) and 
Departamento 
de 
F\'{\i}sica Te\'orica,\\ 
 University of Valencia-CSIC, 
46100 Burjassot, Spain}

\address[cern]{Department of Physics, CERN, 1211 Geneva 23, Switzerland}

\address[uta]{Department of Physics, The University of Texas at Arlington, 
Arlington, TX 
76019, USA}

\address[ddi]{Departamento de Inform\'atica, ETSE, University of Valencia, 
46100 Burjassot, 
Spain}

\cortext[cor1]{Corresponding author}



\begin{abstract}
 We study the impact of a hidden sector beyond the Standard Model, {\em e.g.} a Hidden Valley model, 
on factorial moments and cumulants of multiplicity distributions in 
multiparticle production with a special emphasis on the prospects
for LHC results.
 \end{abstract}

\begin{keyword}
$pp$ interactions at LHC\sep 
models beyond the Standard Model \sep
multihadron correlations \sep
 Hidden Valley models \sep
factorial and cumulant moments\\ 
 \medskip
 {\it Registered preprint number:} arXiv:1510.08738   
\end{keyword}





\maketitle



\section{Introduction}
\label{}

Increasing collision energy at the LHC opens unique opportunities for 
searching signatures of new phenomena beyond the Standard Model (SM). 
On the one hand, some extensions of the SM  
have been suggested by looking for solutions to some pending fundamental 
issues in particle physics. 
This is the case, {\em  e.g.}, of supersymmetry \cite{Haber:1984rc}, and the search for 
supersymmetric particles is one of the fundamental goals of
the LHC. 

On the other hand, there are other scenarios compelled to a lesser extent 
by theoretical arguments, but still motivated by plausible extensions 
of different sectors of the SM, which should be contemplated and its 
phenomenology studied in detail.
 This is for instance the case when a new gauge group (yielding a new type 
of force and a new set of fundamental particles) is added to the theory, 
leading to new bound states with relatively low masses for some values of 
the model parameters. Such scenarios, generically referred as Hidden 
Valley models \cite{Strassler:2006im,Kang:2008ea}, might have remained 
beyond 
observation so far because of an energetic barrier or weak coupling of the 
so-called v-particles to SM particles. Their experimental consequences have 
been already studied having become an objective at the LHC and other
facilities, see {\em e.g.} \cite{Alekhin:2015byh}.
For example, v-hadrons (made of v-quarks) could leave the detector 
undetected, leading to invisible signatures. Alternatively, for some values 
of the parameters of the theory, v-hadrons might decay 
promptly back to SM fermions thereby modifying the parton shower 
hadronizing to final-state particles \cite{Strassler:2008}.

Most signatures of New Physics in colliders are expected to be found 
in hard events, on the transverse plane with respect to
the beams' direction ({\em i.e.} emitting particles with
high transverse momentum $p_{\bot}$), 
where background is much reduced. 
In this work, however, 
we focus on rather diffuse soft
signals in $pp$ inelastic interactions, though expectedly 
tagged by hard decay products and appropriate cuts on events. 
For example, 
 a  non-standard state of matter from a Hidden Sector (HS) 
 might alter 
 particle correlations 
 \cite{SanchisLozano:2008te,Sanchis-Lozano:2014goa}
 which can be measured to a large accuracy at the LHC. 
 Indeed, particle correlations 
 are known to 
 provide a 
 crucial information about 
the 
underlying dynamics of the multiparticle production mechanism
\cite{DeWolf:1995pc,Dremin:2000ep,Manjavidze:2001ni,Kittel:book}  since 
the 
beginning of high-energy 
(cosmic ray)  
physics. In particular, {\em genuine}
 correlations 
 are especially sensitive to variations
 of the features of the partonic cascade leading to final-state particles.

Hadron interactions at high energy are usually considered as resulting 
from collisions of their constituent partons, likely dominated by pairwise 
parton interactions. Such partonic interactions can be hard, leading to 
particles (or jets) with high $p_{\bot}$, or soft, with small transferred 
momentum and large multiplicities. 

 In order to cope with the complexity of multihadron production 
dynamics a 
multi-step scenario, most often a two-step  scenario,  is usually invoked. 
Then the 
resulting final state particle multiplicity distribution and its moments 
are given by the convolution of the distribution of particle emission 
sources such as strings, clusters, fireballs, clans, ladders 
{\em etc.}, 
and their 
decay and/or fragmentation distribution into partons and/or particles. 
Different 
degrees of sophistication can be achieved by introducing various 
phenomenological 
 approaches
 to describe the observed particle 
multiplicity distribution.      

 In particular,
 we will rely on 
 a
 phenonemological approach made in Refs. 
\cite{Dremin:2004ts,Dremin:2011sa} based on the so-called Independent Pair 
Parton Interaction (IPPI) model, in order to study the effects of a new 
physics contribution on the conventional parton cascade. Let us remark 
that the IPPI model does not imply no correlations among the emitted 
particles, but correlations stemming from the distribution probabilities 
describing parton interactions and their convolution, as explained later.

The study of
multiplicity distributions and their properties have traditionally been a cornerstone to understand
soft hadron physics. In this paper, we analyze
 how the multiplicity distributions
of final state hadrons are 
 modified once an extra step 
 of 
 an intermediate unknown state of 
matter is introduced in the 
description of the parton shower. 
 Hereafter we will refer to this approach as the modified IPPI (mIPPI) 
model. Use will be made of the powerful method of 
  the normalised factorial and cumulant moments 
\cite{DeWolf:1995pc,Kittel:book}, allowing to 
extract 
dynamical multiparticle fluctuations and genuine correlations.

\section{Inclusive correlations and factorial moments of 
multiplicity 
distributions}

The study of inclusive particle
correlations in multiparticle production 
can be performed by analyzing 
$n$-particle correlation functions and/or normalised factorial moments 
of multiplicity distributions 
\cite{
DeWolf:1995pc,Dremin:2000ep,Kittel:book}. 
Here we focus on the latter.

The normalized factorial moments of rank $q=2,3,\dots$, are defined as

\begin{equation}\label{eq:Fqdef}
F_q=\frac{\sum_{n}P(n)\ n(n-1)\cdots (n-q+1)}{(\sum_{n}P(n)\ n)^q} \, ,
\end{equation}
where $P(n)$ denotes the probability for $n$ final-state particles 
(charged hadrons).

The factorial moments represent any correlation between the emitted 
particles in events.
To extract the genuine $q$-particle correlations, not reducible to the 
product of the 
lower-order correlations, one uses
the normalised cumulant functions, or cumulants, defined as

\begin{equation}\label{eq:Kqdef}
K_q=F_q-\sum_{r=1}^{q-1}\frac{(q-1)!}{r!(q-r-1)!}K_{q-r}F_r \, .
\end{equation}

The factorial moments and 
cumulants have been extensively
applied  to the analysis of multihadron dynamics in different
types of collisions, from $e^+e^-$ to nucleus-nucleus interactions, 
in a broad
range of energies  \cite{DeWolf:1995pc,Dremin:2000ep,Kittel:book}.

Since $F_q$ and $|K_q|$  grow rapidly as the rank $q$ 
increases, it is convenient to consider the ratio 
\begin{equation}\label{eq:Hq}
H_q=\frac{K_q}{F_q} \, ,
\end{equation}
 which appears in a natural manner as solutions of QCD equations for the 
generating functions of 
multiplicity distributions \cite{Dremin:1993sd}.  

On the other hand, normalized $H_q$ moments are extremely sensitive to the 
details of multiplicity distributions (including experimental cuts on  
 events
 \cite{Ugoccioni:1994zk}) and can be used to distinguish between 
different 
multiparticle production models 
and Monte Carlo generators \cite{Dremin:2000ep,Manjavidze:2001ni}, and 
eventually the contribution of 
a HS as advocated in this paper.


In the following,  
the negative binomial distribution (NBD) which is widely 
used in multiparticle production studies 
\cite{Dremin:2000ep,Kittel:book,ppMultiplicity:Review}, will
 be employed.
 The distribution is given by \cite{Giovannini:1985mz} 
\begin{equation} \label{eq:NBD}
P(n)=\frac{\Gamma(n+k)}{\Gamma(n+1)\Gamma(k)}\biggl(\frac{\langle n \rangle}{k}\biggr)^n
\biggr(1+\frac{\langle n \rangle}{k}\biggr)^{-n-k} \, ,
\end{equation}
where $k^{-1}$ is a parameter which measures how strongly 
 the
 emitted particles
are correlated. One finds that 
\begin{equation} \label{eq:1ok}
\frac{1}{k}=F_2-1 \, .
\end{equation}
The Poisson distribution is obtained in the limit $k\to \infty$ with $F_q= 1$ and $K_q=0$, $\forall q$. 

 In $pp$ interactions, 
one single NBD is found to describe 
satisfactorily
the shape of the charged particle 
multiplicity distribution
at up to 
 several hundreds GeV center-of-mass (c.m.) energy. 
 However, appearance of (shoulderlike) substructures at higher energies 
has been attributed to weighted superposition or convolution of more 
distributions stemming from more than one source or process in 
multiparticle production 
\cite{Fowler:1985fy,Giovannini:1998zb,Giovannini:2003ft}.
 For reviews, see \cite{Kittel:book,ppMultiplicity:Review}. This 
behaviour has been confirmed at LHC energies 
\cite{Zborovsky:2013tla,Ghosh:2013fsa,Adam:2015gka}.
 Thereby one can associate 
the growing complexity with energy of the multiplicity distribution to the 
increasing number of partonic interactions of the colliding particles, 
assuming that every interaction gives rise to a single NBD. This is in 
 fact one of the hypotheses 
 put forward
in \cite{Dremin:2004ts,Dremin:2011sa} that 
we examine in the following section.

\begin{table*}[t]
\setlength{\tabcolsep}{0.6pc}
\caption{Probability distribution of the number of active pairs in proton-proton collisions 
for different TeV energies according to the IPPI model \cite{Dremin:2011sa}.} 
\label{tab:Tab1}

\begin{center}
\begin{tabular}{cccccccc}
\hline $\sqrt{s}$ & $w_1$  & $w_2$ & $w_3$ & $w_4$ & $w_5$ & $w_6$ & $w_7$ \\
\hline $1.8$ TeV & $0.519$ &  $0.269$  & $0.140$ & $0.072$ &  $0.0$ & $0.0$ & $0.0$ \\ 
\hline $7.0$ TeV & $0.504$ &  $0.254$  & $0.128$ & $0.065$ &  $0.033$ & $0.016$ & $0.0$ \\ 
\hline $13$ TeV & $0.5020$ &  $0.2520$  & $0.1265$ & $0.0635$ &  $0.0319$ & $0.0160$ & $0.0080$ \\ 
\hline
\end{tabular}
\end{center}
\end{table*}

\section{Multiparticle production as a multi-step cascade}

 The IPPI model  
\cite{Dremin:2004ts,Dremin:2011sa} was proposed in order 
 reproduce the  moments of multiplicity distributions in $pp$ 
collisions at high energy with minimum adjustable parameters. 
The IPPI  picture corresponds to a simplified 2-step scenario: parton 
binary 
collisions become seeds of independent 
cascades which hadronize ({\em e.g.} via string fragmentation)
to the final-state multiparticle state. 

Moreover, it is assumed that each pair parton interaction gives rise to a
NBD, while the total distribution is ultimately described by means of the  
weighted sum:

\beq\label{eq:conNBD}
P^{(2)}(n) = \sum_{j=1}^{j_{max}}w_j\  \sum_{n_i}\ \prod_{i=1}^{j}\ 
P_{\rm NBD}(n_i,m^{(1)},k^{(1)}) = \sum_{j=1}^{j_{max}}\ w_j\ 
P_{\rm NBD}(n;jm^{(1)},jk^{(1)}) \, ,
\eeq
where $w_j$ denotes the probability for a $j$-pair interaction, 
$m^{(1)}$ and $k^{(1)}$ correspond to the mean 
multiplicity and dispersion for a single pair interaction,
 respectively (for a sake of clarity we explicitly keep in this paper
the superscript $(1)$)\footnote{
Here and in the following, the $k$ parameter in each step is defined as in 
Eq.~(\ref{eq:1ok}), {\em e.g.}  $1/k^{(1)}=F_2^{(1)}-1.$}. Note that
no new adjustable parameters appear in Eq.(\ref{eq:conNBD}) besides the distribution for $j$ binary parton interactions
which can be evaluated if some model is adopted, see {\em e.g.} \cite{Kaidalov:1982xe,Matinyan:1998ja}. 

In the IPPI, the probability for $j$ binary parton interactions per event is simply estimated as $w_j=w_1^j$, 
where $w_1$ refers to a single pair, with the normalization condition $\sum_{j=1}^{j_{max}}w_j=1$. 
In Table~\ref{tab:Tab1} we show the values of $w_j$ up to $w_{max}=7$, 
corresponding to 
$pp$ collisions at the c.m. energy $\sqrt{s}$ 
of 13 TeV  taken from \cite{Dremin:2011sa} (we neglect the 
expectedly slight difference of $w_j$
at c.m. energy between 13 and 14 TeV).  Note that as 
the energy 
increases, more pair parton interactions 
would participate in each event.

Another phenomenological approach based on a QCD-inspired eikonal model
can be found in \cite{Beggio:2013vfa}, leading to similar results for multiciplity distributions and 
factorial moments as the IPPI.


Let us stress that in the current 
 study we do not assume  {\em ab initio} 
any particular type of the
 probability distribution. 
We will keep this general approach in the next section for a 3-step cascade. 
In fact, all
the formulas developed in Appendix apply for any distribution 
at any stage of
the multiparticle production process. Accordingly, a lot of parameters denoted as $F_q^{(p)}$
(where the superscript $p=1,s,h$ will denote different steps of the cascade, 
 {\em vide infra}) 
 encode 
the complexity of the soft hadronic dynamics and hidden production mechanism. Note, however, that
such parameters become fixed once the corresponding probability distributions are adopted.

\begin{table*}[t]
\setlength{\tabcolsep}{0.6pc}
\caption{Normalized factorial moments $F_q^{(s)}$ corresponding to the 
different probability distributions of Table~\ref{tab:Tab1}. Notice that 
moments of rank higher than $w_{max}$ vanish.} \label{tab:Tab2}

\begin{center}
\begin{tabular}{cccccccc}
\hline $\sqrt{s}$ & $F_1^{(s)}$ & $F_2^{(s)}$  & $F_3^{(s)}$ & $F_4^{(s)}$  & $F_5^{(s)}$  & $F_6^{(s)}$  & $F_7^{(s)}$  \\
\hline $1.8$ TeV & $1$ &  $0.72$  & $0.467$ & $0.178$ &  $0.0$ & $0.0$ & $0.0$ \\
\hline $7.0$ TeV & $1$ &  $0.8697$  & $0.8841$ & $0.8353$ &  $0.5979$ & $0.2321$ & $0.0$ \\
\hline $13$ TeV & $1$ &  $0.914$  & $1.050$ & $1.231$ &  $1.256$ & $0.940$ & $0.375$ \\
\hline
\end{tabular}
\end{center}
\end{table*}

\subsection{Two-step cascade}

 One can rewrite  Eq.(\ref{eq:conNBD}) of the independent superposition of 
parton 
pair interactions 
in $pp$ collisions 
 for  arbitrary particle production distributions and  sources:
\begin{equation}\label{eq:ippi-2}
P^{(2)}(n)\ =\ \sum_{N_s}\ P(N_s)\ \sum_{n_i}\ \prod_{i=1}^{N_s}\ 
P^{(1)}(n_i) \, .
\end{equation}
Here $n$ and $N_s$ denote the number of (charged) particles and sources, 
respectively
\footnote{To compare with the experimental data and
to get the particle multiplicity with the same charge, 
the multiplicity was divided by 
 two in \cite{Dremin:2004ts,Dremin:2011sa}. 
Also note that the number of sources $N_s$ in Eq.({\ref{eq:ippi-2})} corresponds
to the number of parton pair collisions $j$ in Eq.(\ref{eq:conNBD}).}. In the 
notation used here,
$P(N_s)$ stands now for the distribution of (fragmenting string) sources, equivalent to the parton pair interaction
distribution $w_j$. Correspondingly, the average multiplicity
can be written as $\langle n \rangle = \langle N_s \rangle\ m^{(1)}$ according to a 2-step
description of multiparticle production.

On the other hand, the authors of \cite{Dremin:2004ts,Dremin:2011sa} benefit from a dramatic reduction of
free parameters when assuming a weighted superposition of NBDs with shifted parameters, as can be seen
 in Eq.(\ref{eq:conNBD}). In addition, since $m^{(1)}$ 
 should be the same for any value of the rank $q$, only $k^{(1)}$ remains 
 a free parameter ($w_{max}$ was determined using a particular model). 
Remarkably, in 
the current analysis, $m^{(1)}$
cancels out in the expressions for the scaled factorial 
moments and cumulants.

 In order to make a comparison of the results of the current study and 
those from 
\cite{Dremin:2004ts,Dremin:2011sa}, below we assume that all 
$P^{(1)}(n_i)$  are NBDs. 
Moreover, $P(N_s)$ and $w_j$ distributions can be formally identified. 
 The values of $F_q^{(s)}$ up to $q=7$ are 
 given 
in 
 Table~\ref{tab:Tab2}
(higher rank moments vanish for $w_{max}=7$). 
 Let us 
 stress that they 
 do  
 {\bf not} 
 represent a  
 NBD.

Upon integration of the inclusive correlation functions in the
central rapidity region \cite{Sanchis-Lozano:2014goa}, the $F_q^{(2)}$ moments can be written in terms of
the moments of the subprocesses of the cascade;
 for example, the factorial moments of rank two read
\begin{equation}\label{eq:F2twostep}
F_2^{(2)}=F_2^{(s)}+\frac{F_2^{(1)}}{\langle N_s \rangle} \, .
\end{equation}
Here, $F_2^{(s)}=\langle N_s(N_s-1) \rangle/\langle N_s \rangle^2$ and 
$F_2^{(1)}=\langle n_1(n_1-1) \rangle/\langle n_1 \rangle^2$, and 
$\langle n_1 \rangle=m^{(1)}$ stands for the average particle multiplicity per single cascade.

 The computation of higher rank $F_q^{(2)}$ moments becomes extremely 
involved at large $q$. Therefore, 
we have written a Prolog code (see Appendix)
which provides the expressions $F_q^{(p)}$ for any value of the 
rank $q$ and any number of steps $p$ in the cascade, depending on the 
 computer capacity available.

As shown below,  
we are able to reproduce (up to the percent level) the $H_q^{(2)}$ moments 
\footnote{As in the case of $F_q^{(p)}$ moments, here the superindex $p$ 
in
$H_q^{(p)}$ indicates the number of steps in 
the cascade: a two-step 
 conventional cacade with $p=2$, 
and the three-step cascade with $p=3$ once a HS is 
 included.}
using the same values and assumptions as in Refs. 
\cite{Dremin:2004ts,Dremin:2011sa}. 
This accordance 
 suggests to proceed  
further in the approach given 
here by 
incorporating a new step in the parton cascade
following the mIPPI scheme.

Interestingly, the values of $H_q^{(2)}$ can become 
quite small (down to a 
decimal order, even approaching zero for certain 
values
of $q$) while the factorial moments $F_q^{(2)}$ grow fast with $q$. 
Actually 
there is a delicate balance in the cancellations
of Eq.(\ref{eq:Kqdef}) which can be altered when the characteristics of the parton cascade vary. Such a sensitivity could
be of utility in the search for new phenomena in hadron collisions, as 
it is advocated here.

\begin{figure*}[t]
\begin{center}
\includegraphics[scale=1.0]{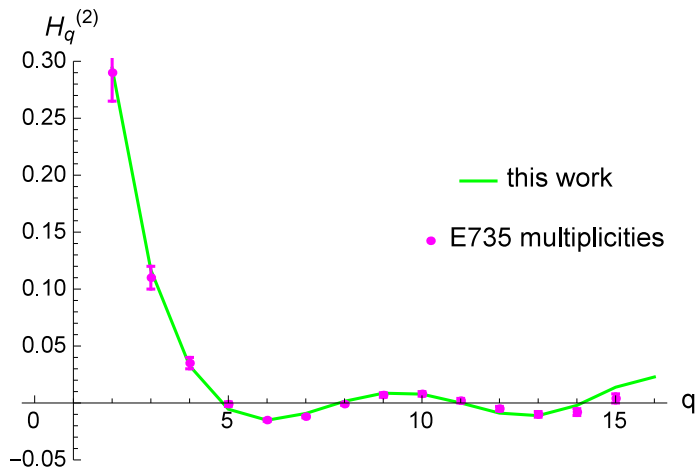}
\end{center}
\caption{$H_q^{(2)}$ moments up to $q=16$ in $p{\bar p}$ collisions at 1.8 TeV 
obtained in this work using expressions for a 2-step cascade and the same 
parameters 
 as in 
 \cite{Dremin:2004ts,Dremin:2011sa}.
Very good agreement is found with the results from \cite{Dremin:2004ts,Dremin:2011sa} 
 and with the 
 calculations based on the 
 multiplicity
 measurements 
 \cite{E735};
 the latter 
 shown by circles with error bars.}  
\label{fig:Fig1}
\end{figure*}

\subsection{Three-step cascade}

Let us now include an extra step in the cascade to simulate a hypothetical 
new stage of matter associated to a HS. The resulting multiplicity in a 3-step process 
should obey the following distribution:
\begin{equation}\label{eq:ippi-3}
P^{(3)}(n)\ =\ \sum_{N_s}\ P(N_s)\ \sum_{n_j}\ \prod_{j=1}^{N_s}\ 
P^{(2)}(n_j) \, .
\end{equation}
 where $P^{(2)}$  is here defined  as
\begin{equation}\label{eq:ippi-2h}
P^{(2)}(n)\ =\ \sum_{N_h}\ P(N_h)\ \sum_{n_i}\ \prod_{i=1}^{N_h}\ 
P^{(1)}(n_i) \, ,
\end{equation}
with $N_h$ denoting the number of active hidden sources in a collision. 
 In what follows, 
for the sake of simplicity we  assume that $P(N_h)$ follows a Poisson 
distribution, {\em i.e.} independent production of hidden sources 
resulting from 
binary parton interactions.

In other words, the probability distribution of parton interactions 
remains the same as in the conventional cascade 
(being already adjusted to reproduce experimental data in $pp$ collisions) 
while 
one adds  another step subsequent to
the initial binary parton interaction.

 Then, proceeding in the same way as in  
in the previous section,
 one gets the $F_q^{(3)}$ moments in terms of the 
 multiplicity moments from the different cascade steps. 
 For example, the second-rank factorial moment $F_2^{(3)}$ reads
 
\begin{equation}\label{eq:F2threestep}
F_2^{(3)}=F_2^{(s)}+\frac{F_2^{(h)}}{\langle N_s \rangle}+
\frac{F_2^{(1)}}{\langle N_h \rangle} \, ,
\end{equation}
 where $\langle N_h \rangle$ and $F_2^{(h)}$ stand for
the mean number and scaled moment of the hidden source distribution, 
 respectively. In the Appendix
 the expressions for moments of rank up to $q=6$ 
(showing increasing complexity) are provided.

As already commented, the computation of $F_q^{(p)}$ becomes 
especially hard for high
$q$ values and the above-mentioned Prolog code is used
to obtain further factorial moments and cumulants. For example,
$F_{16}^{(3)}$ contains about 100,000 terms with some coefficients
of numerical order $10^9$. Needless to say again, we have checked carefully
the numerical stability of the computation.

 We
have also checked the first $F_q^{(p)}$ moments, $p=2$ and $3$,  up 
to $q=8$ obtained with the Prolog code to those  
computed {\em by hand} and shown in  Appendix up to $q=6$.\footnote{In 
ref.~\cite{Sanchis-Lozano:2014goa} we already presented some 
of these expressions for low values of $q$ in
two-, three- and even four-step scenarios. In this paper, however, we do 
 limit ourselves to a three-step scenario leaving the 
four-step scenario to be considered elsewhere.
Beware also of the notation change 
of superscripts with respect to the present study.}

\begin{figure*}[t]
\begin{center}
\includegraphics[scale=0.57]{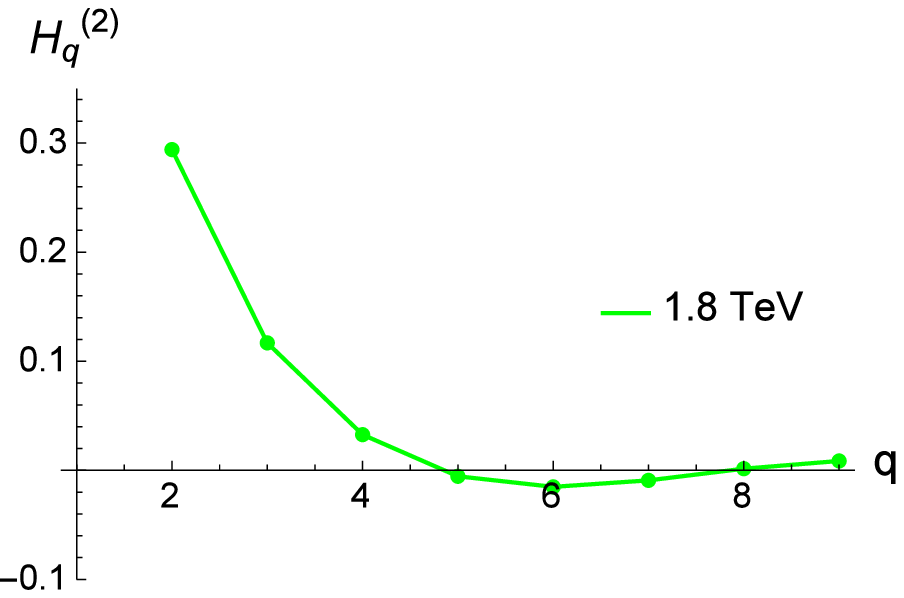}
\includegraphics[scale=0.57]{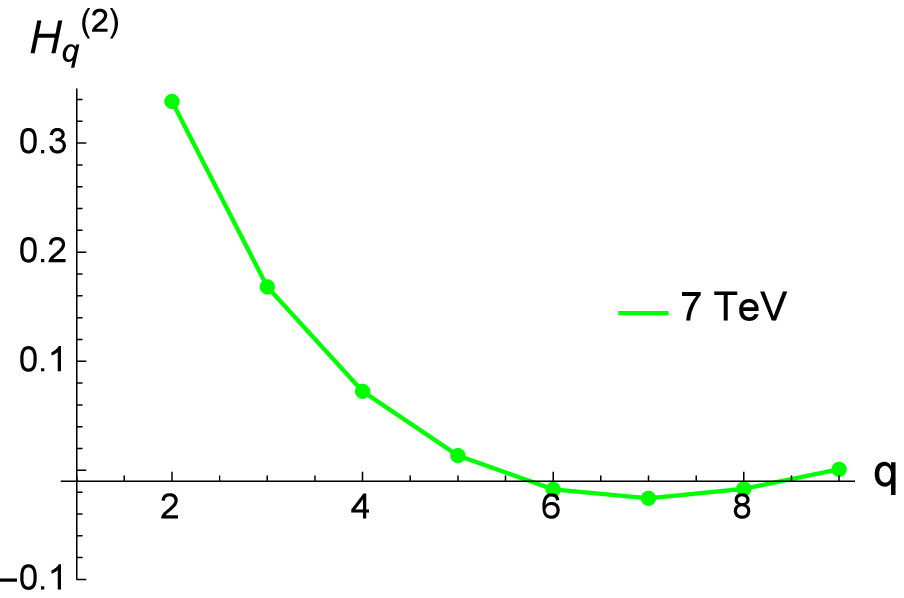}
\includegraphics[scale=0.57]{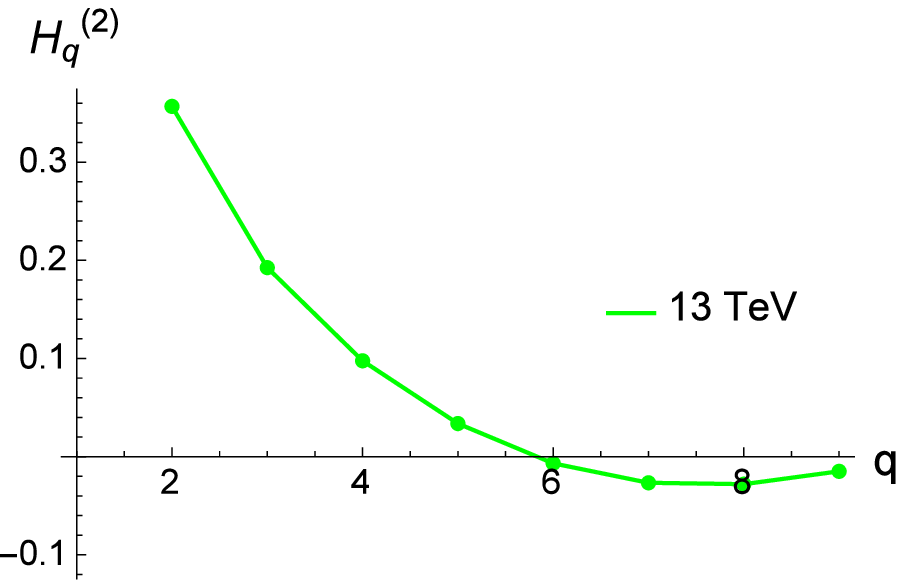}
\end{center}
\caption{$H_q^{(2)}$ moments calculated up to $q=8$ in $pp$ 
 collisions at $\sqrt{s}=1.8$~TeV, 
 7 TeV 
 and 13 TeV 
 (from left to right) 
  according to a 
conventional (2-step) cascade. The interpolating line is plotted
to guide the eye.
 The crossing point (and first minimum) moves to higher $q$ values as the 
 collision energy increases.}  
\label{fig:Fig2}
\end{figure*}


 \section{$H_q$-moment oscillations as a function of the rank $q$}

QCD next-to-leading order
calculations \cite{Dremin:1994wm,Buican:2003jx} predict that the ratios $H_q$
 defined in Eq.(\ref{eq:Hq}) oscillate as a function of the rank $q$, 
crossing the $q$-axis and
becoming negative  with a minimum at
\begin{equation} \label{eq:qmin}
q_{min} \approx \frac{24}{11}\frac{1}{\gamma_0}+\frac{1}{2}+{\cal 
O}(\gamma_0) \, ,
\end{equation}
where $\gamma_0=(6\alpha_s/\pi)^{1/2}$ denotes the anomalous dimension at 
lowest order; for review, see \cite{Dremin:2000ep,Khoze:1996dn}. 
 At LEP energies, it turns out
that  $q_{min} \approx 5$ shifting to larger values at higher 
energies. 
This prediction has been tested against experimental data and found to be 
observed not only in $e^+e^-$ collisions \cite{Abe:1996vs,Achard:2001ut} 
but also in 
a variety of 
colliding particles and energies, including $pp$, $pA$ and $AA$ collisions 
{\cite{Dremin:1997hk}}. 

It is relevant to emphasize here that in case of a single NBD, the 
cumulants $H_q$ are always positive (hence no oscillations appear) and 
monotonically decreasing as a function of $q$, in clear disagreement with 
the QCD predictions and experimental data \cite{Dremin:2000ep}.  
   The study of factorial moments and cumulants also reveal difficulties 
that the NBD faces to describe multiparticle production in 
full phase space and in its small
intervals \cite{Kittel:book,Sarkisian:2000ux,Abbiendi:1999ki,Abbiendi:2001bu}.

In Fig.~\ref{fig:Fig1}  we plot the values of the $H_q^{(2)}$ moments 
($q=2$ to 
16) for 
$\sqrt{s}=1.8$ TeV multiplicity data, obtained through 
Eqs.(\ref{eq:Kqdef}) and (\ref{eq:Hq}) from the expressions of $F_q^{(2)}$ 
(a 2-step cascade). We fix
the parameters for the plot alike it is done in Ref.~\cite{Dremin:2004ts}, 
{\em i.e.}, assuming NBDs for all binary parton collisions  
with $k^{(1)}=4.4$,  and $P(N_s)$ (equivalent to the $w$ distribution) 
from Table~\ref{tab:Tab1}. The overall agreement 
with the results of 
Refs. \cite{Dremin:2004ts,Dremin:2011sa} 
and experimental data \cite{E735} is very good.

 One can see the 
 two minima in the Fig.~\ref{fig:Fig1}. As later interpreted, 
this oscillatory 
pattern 
(which 
 seems to 
 continue for even higher ranks) is due to the
fact that the probability distribution for the number of sources $P(N_s)$ 
(equivalently, the distribution for 
the number of parton pair
collisions) does {\bf not} follow a NBD. In case the 
distributions 
are all negative binomial, the resulting distribution turns out 
to be of the 
NBD type too and no oscillation pattern for $H_q^{(2)}$ shows 
up.

In Fig.~\ref{fig:Fig2} 
the $H_q^{(2)}$ moments 
obtained in the current study
are shown 
 for different 
$pp$ collision
c.m. energies
as a function of the rank 
$q$ 
being limited to 
the first miminum. Namely, the 
moments 
$H_q^{(2)}$ are plotted for $\sqrt{s}=1.8$ TeV, 
7 TeV and 13 TeV. The points were evaluated 
computing first the values of the $F_q^{(s)}$ moments corresponding
to the 
source (or binary parton) probability distribution at different energies, 
shown in 
Table~\ref{tab:Tab2}. We set $k^{(1)}=4$ as an input in the calculations here 
similarly to 
the value of this parameter used in \cite{Dremin:2004ts,Dremin:2011sa}. 
One can see indeed that the minimum moves to the right as the $pp$ collision energy 
increases, as expected. 

 The good
 agreement with the 
 measurements 
 shown 
 in Fig.~\ref{fig:Fig1} and the 
 expectations with the collision energy shown in a set of plots of 
Fig.~\ref{fig:Fig2} 
 suggests the further
 introduction of 
 a new step in the cascade 
to be interpreted as a HS, thereby
studying the eventual variation of the crossing points/minima and the 
amplitude of the $H_q$ oscillations.

\begin{figure*}[t]
\begin{center}
\includegraphics[scale=1.]{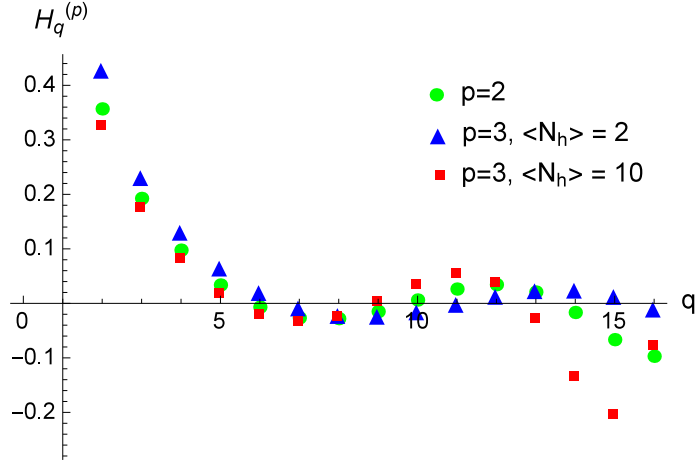}
\end{center}
\caption{Predictions for $H_q^{(p)}$ moments as a function of 
 the rank $q$  for $pp$ collisions at $\sqrt{s}=13$~TeV. The 
 circles 
correspond to a 
conventional 2-step cascade ($p=2$) from extrapolation at lower energies 
using 
 the IPPI model. The 
 triangles and the  squares correspond to a 3-step cascade ($p=3$) 
using the mIPPI model (this work) with 
the number of hidden sources 
$\langle N_h \rangle =2$
and $\langle N_h \rangle =10$, respectively. 
A different pattern in the amplitude of the oscillations
at high $q$ values can be clearly observed.}
\label{fig:Fig3}
\end{figure*}

\section{HS-cascade versus a conventional cascade}

\subsection{Shift of the first minimum of $H_q$ as a function of $q$}

As explained above, the behaviour of the first minimum of $H_q$ with 
the c.m. energy in $pp$ collisions 
is well 
predicted.
Let us now examine how
this behaviour can be modified in a 3-step scenario under different 
assumptions.

In Fig.~\ref{fig:Fig3}, the
 three sets of points corresponding to different scenarios at 
$pp$ collisions at $\sqrt{s}=
13$~TeV  
 are shown. The 
 circles correspond to a conventional cascade, while 
 the
 triangles and 
 squares
 correspond to an extra step in the mIPPI model setting
$\langle N_h  \rangle=2$ and $\langle N_h  \rangle =10$, 
respectively.
One can see that the  crossing point (and minimum) moves by about one unit 
to the left
for
 $\langle N_h  \rangle =10$, and by the same amount 
to the right 
for 
 $\langle N_h  \rangle =2$ 
 compared to the case of a conventional cascade.  
Such an altered behaviour could become a hint of a HS affecting the 
parton evolution in multiparticle production, deserving a more detailed study.

\begin{figure*}[t]
\begin{center}
\includegraphics[scale=0.86]{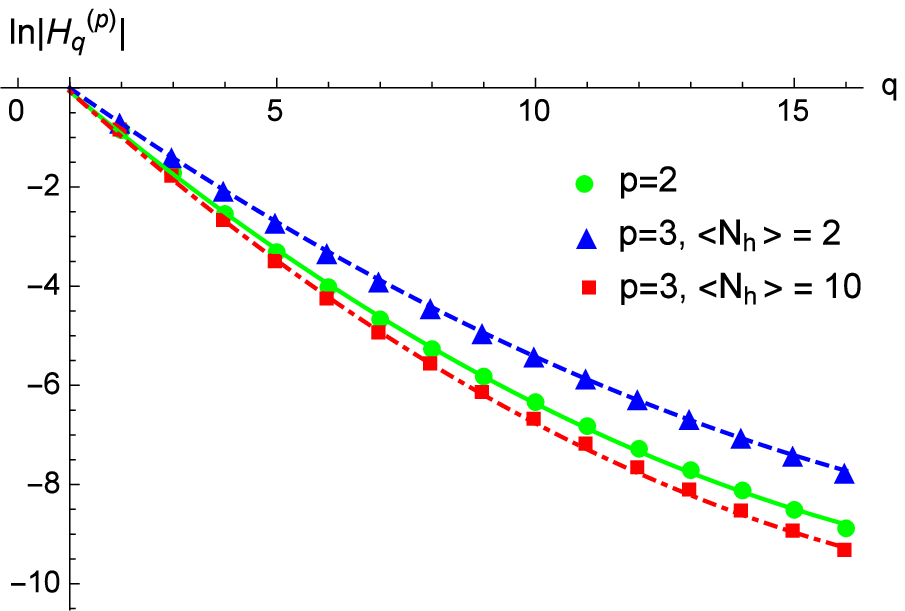}
\includegraphics[scale=0.62]{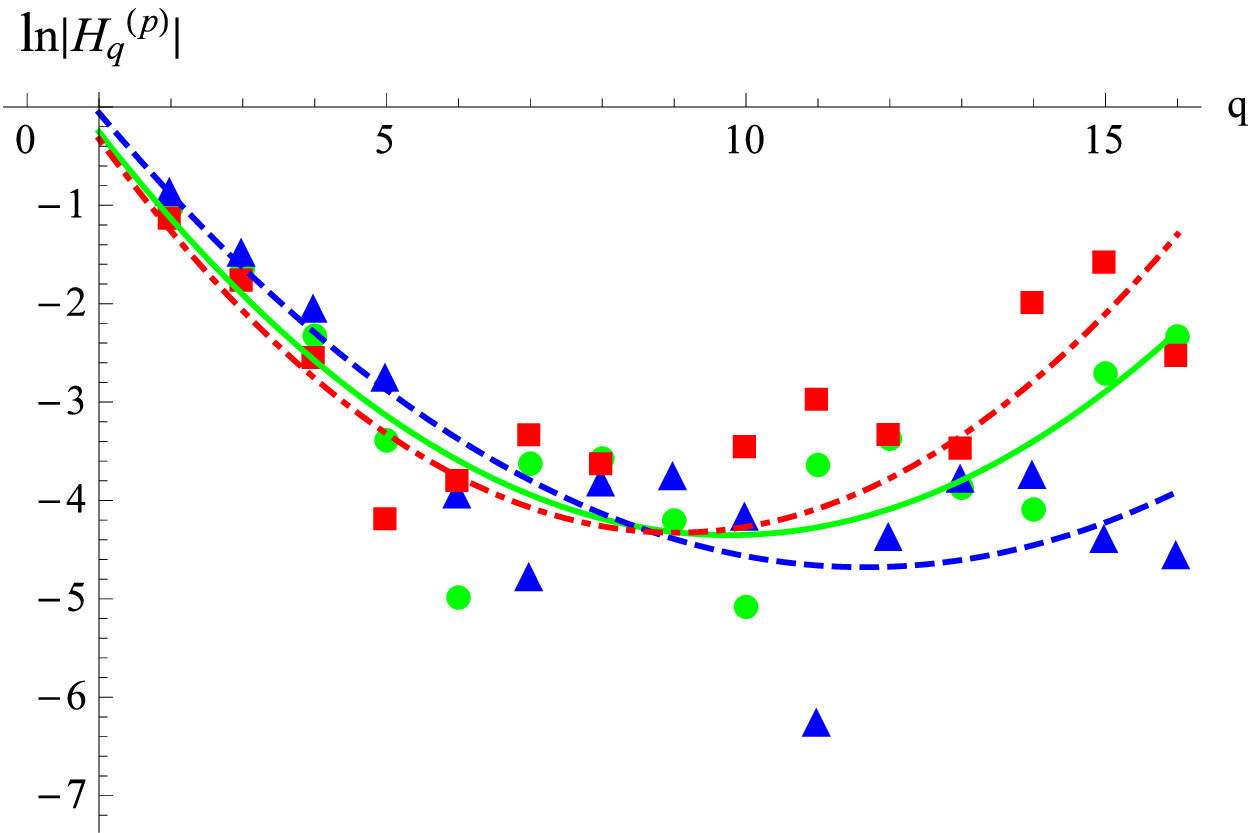}
\end{center}
\caption{Values of $\ln|H_q^{(p)}|$ at $\sqrt{s}=13$~TeV versus 
 $q$ for a 2-step scenario ($p=2$,
 circles) and 
a 3-step scenario 
($p=3$) 
 with $\langle N_h \rangle =2$ 
 (triangles) and 
$\langle  N_h \rangle=10$ 
 (squares).
Solid, dashed and dot-dashed lines
correspond to parabolic fits, respectively.
 Left panel:
superposition of NBDs with $k^{(1)}=4$ and $k^{(s)}=k^{(h)}=10$ 
as reference values. 
Right panel: $P(N_s)$
 incorporates
 the values of Table~\ref{tab:Tab2} 
 (not the NBD case). 
Notice that there are no oscillations when {\em all} the distributions 
of the convolution are of the NBD type.}  
\label{fig:Fig4}
\end{figure*}

\subsection{Analysis of the $H_q$ oscillation amplitude}

Next let us examine the amplitude of the $H_q$ 
oscillations as a 
function of the rank $q$, and its dependence on the parameters used in 
the mIPPI model
 as can be seen from in Fig.~\ref{fig:Fig3}. 
 %
 
Depending on the number of hidden sources two different behaviours of 
the oscillation pattern of $H_q^{(3)}$ moments can be distinguished:

\begin{itemize}

\item For a small number of hidden sources, the oscillation
amplitude becomes appreciably dumped for high $q$ values
as compared to a conventional (2-step) cascade.
 
\item For a large number of hidden sources, the oscillation
amplitude is considerably larger for high $q$ values
as compared to a conventional (2-step) cascade.

\end{itemize}

These conclusions are indeed confirmed in Fig.~\ref{fig:Fig4} where 
 the values of 
$\ln|H_q^{(p)}|$ are 
plotted against $q$ for different scenarios depending on the 
type of the distributions used as indicated.
 The calculated points are shown 
together with the parabolic fits to them. 
One can see that the behaviour of the fitted curves is very different for 
different scenarios 
especially at large $q$ values, which could thereby be relevant to detect
a new physics effect according to the study presented here.

The fitted curves pass through the points
in the left panel, whereas the points
scatter around 
the curves in the right panel. This means that no 
oscillations appear whenever all the distributions in 
the superposition of Eqs.(\ref{eq:ippi-3}) and (\ref{eq:ippi-2h}), 
including 
$P(N_s)$, are of the NBD type. As already commented in Section 4, this 
behaviour can
be easily understood in the mIPPI model
inasmuch the convolution of NBDs in Eq.(\ref{eq:ippi-3})
leads again to a NBD.
Conversely, the oscillation pattern in the right panel emerges as a 
consequence of $P(N_s)$ not being a NBD. 

\section{Summary and final remarks}

In this work we advocate that a new stage of matter, stemming from a 
 hidden sector beyond the SM on top of the conventional partonic cascade, can 
be observed in multiparticle production in 
 $pp$ collisions at 
the LHC.
 This would result 
on some features 
of final-state particle correlations measured using the technique of 
factorial 
and cumulant moments of 
multiplicity distributions.

 Within the  modified Independent Parton Pair Interaction  (mIPPI) model 
(with an extra step in addition
to the 
conventional 2-step IPPI model {\cite{Dremin:2004ts,Dremin:2011sa}}), 
 the effect of a HS on the cumulant-to-factorial moment ratio $H_q$  
 of 
the multiplicity 
distributions of final-state 
particles strongly depends on the number of the hidden sources. A large 
 (small) number of the sources 
 would lead to an
enhancement (softening) of the 
oscillation amplitude 
 at high 
$q$ 
values. 
Moreover, the crossing of the $q$-axis and the minimum of the 
$H_q$-moments 
interpolating curve
shifts to smaller (larger) $q$ values for a large (few) number of hidden 
sources. 

We have provided new expressions for the scaled factorial moments 
$F_q^{(p)}$ of the two-  and three-step  cascades ($p=2$ and 3, 
respectively) 
not given in the literature to our knowledge. 
Some of them (up to $q=6$) are explicitly written in  Appendix. Higher 
rank factorial
moments can be computed by means of the
Prolog code developed for this work, leading to very long and complicated formulas.
 We have carefully checked the correctness of our code by comparing the 
computed expressions to 
 {\em by-hand} calculations up to $q=8$. The numerical stability at large 
$q$ was also tested
by choosing, {\em e.g.}, a Poisson distribution for all intermediate probability distributions, and
checking that the resulting values allign in the
$\ln|H_q^{(p)}|$ plot as a function of $q$ (see Fig.~\ref{fig:Fig4}).



To conclude, we have studied the phenomenological consequences of HS 
physics in multiparticle production
which could be useful at LHC experiments, likely requiring a 
low-luminosity run to reduce
pile-up as much as possible. 
 Since both --conventional and HS -- processes would be present in 
the collected sample of events, specific cuts, such as high 
multiplicity, flavour tagging, high-$p_{\bot}$ leptons, missing energy, 
{\em etc.}, are suggested  to be applied  to 
enrich the 
 signatures of 
 new physics. 

\section*{Acknowledgments}

This work has been partially supported by MINECO under grants 
FPA2011-23596 and FPA2014-54459-P, and Generalitat Valenciana under grant 
PROMETEOII/2014/049.
 One of us (M.A.S.L.) acknowledges support from IFIC under grant 
SEV-2014-0398 of the ``Centro de Excelencia Severo Ochoa'' Programme.

\newpage



\appendix

\section{Factorial moments in 2- and 3-step scenarios}


\subsection{Prolog code}

 The structure of the problem suggests the use of declarative programming 
or functional
programming. Finally the chosen language was Prolog \cite{prolog}.\\

The program we have used to compute the factorial moments consists of four parts \cite{prolog:2015}:
\begin{itemize}
\item[1)] A predicate which generates all possible topologies without repetitions from a given
number of final particles and a given number of disintegration steps.
\item[2)] A recursive predicate that counts the number of occurrences for each topology.
\item[3)] A predicate that groups all topologies generated under a common formula, adding all
occurrences per formula.
\item[4)] A predicate that translates formulas generated as a latex file and as a Mathematica
file.
\end{itemize}
The final result can be incorporated into a latex document, or can be incorporated
directly in Mathematica to use the results in calculations and generate 
 the corresponding
graphs.

\subsection{Modelled expessions for factorial moments}

 In this Appendix we give the expressions for the 
 factorial moments 
 $F_q^{(p)}$ up to $q=6$. 
 The superscript $p=2$ and $3$  denotes  a 2- and 3-step scenario, 
 respectively.
 The expressions for higher rank moments 
 are too long to be reproduced here but can be obtained 
from the Prolog code developed for this work. \\

\noindent
{\bf 2-step cascade}
\[
F_2^{(2)}=F_2^{(s)}+\frac{F_2^{(1)}}{\langle N_s \rangle}\ \ \ ,\ \ \ 
F_3^{(2)}=F_3^{(s)}+3\frac{F_2^{(s)}}{\langle N_s \rangle}F_2^{(1)}
+\frac{F_3^{(1)}}{\langle N_s \rangle^2}
\]

\[
F_4^{(2)}=F_4^{(s)}+\frac{F_4^{(1)}}{\langle N_s \rangle^3}+
6\frac{F_3^{(s)}F_2^{(1)}}{\langle N_s \rangle}+
4\frac{F_2^{(s)}F_3^{(1)}}{\langle N_s \rangle^2}+
3\frac{F_2^{(s)}F_2^{(1)2}}{\langle N_s \rangle^2}
\]

\[
F_5^{(2)} = F_5^{(s)}+
\frac{F_5^{(1)}}{\langle N_s \rangle^4}+
10\frac{F_4^{(s)}F_2^{(1)}}{\langle N_s \rangle}+
10\frac{F_3^{(s)}F_3^{(1)}}{\langle N_s \rangle^2}+ 
15\frac{F_3^{(s)}F_2^{(1)2}}{\langle N_s\rangle^2}+ 
5\frac{F_2^{(s)}F_4^{(1)}}{\langle N_s \rangle^3}+ 
10\frac{F_2^{(s)}F_3^{(1)}F_2^{(1)}}{\langle N_s \rangle^3}     
\]

\begin{eqnarray*}
F_6^{(2)} &=& F_6^{(s)}+
\frac{F_6^{(1)}}{\langle N_s \rangle^5}+ 
15\frac{F_5^{(s)}F_2^{(1)}}{\langle N_s \rangle}
+20\frac{F_4^{(s)}F_3^{(1)}}{\langle N_s \rangle^2}+ 
15\frac{F_3^{(s)}F_4^{(1)}}{\langle N_s \rangle^3}+
6\frac{F_2^{(s)}F_5^{(1)}}{\langle N_s \rangle^4}+ 
\nonumber \\
&& 
15\frac{F_2^{(s)}F_4^{(1)}F_2^{(1)}}{\langle N_s \rangle^4}+
10\frac{F_2^{(s)}F_3^{(1)2}}{\langle N_s \rangle^4}+ 
60\frac{F_3^{(s)}F_3^{(1)}F_2^{(1)}}{\langle N_s \rangle^3}+
45\frac{F_4^{(s)}F_2^{(1)2}}{\langle N_s \rangle^2}+
15\frac{F_3^{(s)}F_2^{(1)3}}{\langle N_s \rangle^3} \nonumber
\end{eqnarray*}

\vskip 0.2cm

\noindent
{\bf 3-step cascade}

\[
F_2^{(3)}=F_2^{(s)}+\frac{F_2^{(h)}}{\langle N_s \rangle}+
\frac{F_2^{(1)}}{\langle N_h \rangle}\ \ \, \ \ \
F_3^{(3)}=F_3^{(s)}+\frac{F_3^{(h)}}{\langle N_s \rangle^2}+
3\biggl[\frac{F_2^{(s)}F_2^{(h)}}{\langle N_s \rangle}+\frac{F_2^{(s)}
F_2^{(1)}}{\langle N_h \rangle}+
\frac{F_2^{(h)}F_2^{(1)}}{\langle N_s \rangle \langle N_h \rangle}
\biggr]+\frac{F_3^{(1)}}{\langle N_h \rangle^2}
\]

\[
F_4^{(3)}=F_4^{(s)}+\frac{F_4^{(h)}}{\langle N_s \rangle^3}+ 
\frac{F_4^{(1)}}{\langle N_h \rangle^3}+ 
6\biggl[
\frac{F_3^{(s)}F_2^{(h)}}{\langle N_s \rangle}+
\frac{F_3^{(s)}F_2^{(1)}}{\langle N_h \rangle}+
\frac{F_3^{(h)}F_2^{(1)}}{\langle N_s \rangle^2 \langle N_h \rangle}
\biggr]+ 
4\biggl[\frac{F_2^{(s)}F_3^{(h)}}{\langle N_s \rangle^2}+
\frac{F_2^{(s)}F_3^{(1)}}{\langle N_h \rangle^2}+
\frac{F_2^{(h)}F_3^{(1)}}{\langle N_s \rangle \langle N_h \rangle^2}
\biggr]+ 
\]
\[
3\biggl[\frac{F_2^{(s)}F_2^{(h)2}}{\langle N_s \rangle^2}+
\frac{F_2^{(s)}F_2^{(1)2}}{\langle N_h \rangle^2}+
\frac{F_2^{(h)}F_2^{(1)2}}{\langle N_s \rangle \langle N_h \rangle^2}\biggr]+ 
18\frac{F_2^{(s)}F_2^{(h)}F_2^{(1)}}{\langle N_s \rangle \langle N_h \rangle}
\]

\begin{eqnarray}\label{eq:F5u}
F_5^{(3)} &=& F_5^{(s)}+\frac{F_5^{(h)}}{\langle N_s \rangle^4}+
\frac{F_5^{(1)}}{\langle N_h \rangle^4}+ 
10\biggl[\frac{F_4^{(s)}F_2^{(h)}}{\langle N_s \rangle}+
\frac{F_4^{(s)}F_2^{(1)}}{\langle N_h \rangle}+
\frac{F_4^{(h)}F_2^{(1)}}{\langle N_s \rangle^3 \langle N_h \rangle}\biggl]+
\nonumber \\
&& 5\biggl[\frac{F_2^{(s)}F_4^{(h)}}{\langle N_s \rangle^3}+
\frac{F_2^{(s)}F_4^{(1)}}{\langle N_h \rangle^3}+
\frac{F_2^{(h)}F_4^{(1)}}{\langle N_s \rangle \langle N_h \rangle^3}
\biggr]
+10\biggl[\frac{F_3^{(s)}F_3^{(1)}}{\langle N_h \rangle^2}+
\frac{F_3^{(s)}F_3^{(h)}}{\langle N_s \rangle^2}+
\frac{F_3^{(h)}F_3^{(1)}}{\langle N_s \rangle^2\langle N_h \rangle^2}\biggl]+ \nonumber \\
&&
15\biggl[
\frac{F_3^{(s)}F_2^{(h)2}}{\langle N_s \rangle^2}+
\frac{F_3^{(s)}F_2^{(1)2}}{\langle N_h \rangle^2}+
\frac{F_3^{(h)}F_2^{(1)2}}{\langle N_s \rangle^2 \langle N_h\rangle^2}
\biggl]+
10\biggl[
\frac{F_2^{(h)}F_3^{(1)}F_2^{(1)}}{\langle N_s \rangle \langle N_h \rangle^3}+
\frac{F_2^{(s)}F_3^{(1)}F_2^{(1)}}{\langle N_h \rangle^3}+ \nonumber \\
&&
4\frac{F_2^{(s)}F_3^{(h)}F_2^{(1)}}{\langle N_s \rangle^2 \langle N_h \rangle}+
3\frac{F_2^{(s)}F_2^{(h)}F_3^{(1)}}{\langle N_s \rangle \langle N_h \rangle^2}+ \frac{F_2^{(s)}F_3^{(h)}F_2^{(h)}}{\langle N_s \rangle^3}+
6\frac{F_3^{(s)}F_2^{(h)}F_2^{(1)}}{\langle N_s \rangle \langle N_h \rangle}
\biggl]+ \nonumber \\
&&
30\biggl[
\frac{F_2^{(s)}F_2^{(h)2}F_2^{(1)}}{\langle N_s \rangle^2 \langle N_h \rangle}+
1.5\frac{F_2^{(s)}F_2^{(h)}F_2^{(1)2}}{\langle N_s \rangle \langle N_h \rangle^2}
\biggr] \nonumber
\end{eqnarray}

\begin{eqnarray}\label{eq:F6u}
F_6^{(3)} &=& F_6^{(s)}+\frac{F_6^{(h)}}{\langle N_s \rangle^5}+
\frac{F_6^{(1)}}{\langle N_h \rangle^5}+ \nonumber \\
&&
6\biggl[\frac{F_2^{(s)}F_5^{(h)}}{\langle N_s \rangle^4}+
\frac{F_2^{(s)}F_5^{(1)}}{\langle N_h \rangle^4}+\frac{F_2^{(h)}F_5^{(1)}}
{\langle N_s \rangle \langle N_h \rangle^4}\biggr]+
15\biggl[\frac{F_3^{(s)}F_4^{(1)}}{\langle N_h \rangle^3}+\frac{F_3^{(s)}F_4^{(h)}}{\langle N_s \rangle^3}+ 
\frac{F_3^{(h)}F_4^{(1)}}{\langle N_s \rangle^2\langle N_h \rangle^3}\biggr]+ \nonumber \\
&&
20\biggl[\frac{F_4^{(s)}F_3^{(h)}}{\langle N_s \rangle^2}+
\frac{F_4^{(s)}F_3^{(1)}}{\langle N_h \rangle^2}+
\frac{F_4^{(h)}F_3^{(1)}}{\langle N_s \rangle^3 \langle N_h \rangle^2}\biggr]+
15\biggl[\frac{F_5^{(s)}F_2^{(h)}}{\langle N_s \rangle}+
\frac{F_5^{(s)}F_2^{(1)}}{\langle N_h \rangle}
+\frac{F_5^{(h)}F_2^{(1)}}{\langle N_s \rangle^4 \langle N_h \rangle} \biggr]+
\nonumber \\
&&
15\biggl[
5\frac{F_2^{(s)}F_4^{(h)}F_2^{(1)}}
{\langle N_s \rangle^3 \langle N_h \rangle}+
\frac{F_2^{(s)}F_4^{(h)}F_2^{(h)}}
{\langle N_s \rangle^4 \rangle}+
3\frac{F_2^{(s)}F_2^{(h)}F_4^{(1)}}
{\langle N_s \rangle \langle N_h \rangle^3}+
\frac{F_2^{(s)}F_4^{(1)}F_2^{(1)}}{\langle N_h \rangle^4}+
\frac{F_2^{(h)}F_4^{(1)}F_2^{(1)}}{\langle N_s \rangle \langle N_h \rangle^4}
\biggr]+ \nonumber \\
&& 10\biggl[\frac{F_2^{(s)}F_3^{(h)2}}{\langle N_s \rangle^4}+
8\frac{F_2^{(s)}F_3^{(h)}F_3^{(1)}}{\langle N_s \rangle^2 \langle N_h \rangle^2}+
\frac{F_2^{(s)}F_3^{(1)2}}{\langle N_h \rangle^4}+
\frac{F_2^{(h)}F_3^{(1)2}}{\langle N_s \rangle \langle N_h \rangle^4}\biggr]+ \nonumber \\
&&60\biggl[
\frac{F_3^{(s)}F_3^{(h)}F_2^{(h)}}{\langle N_s \rangle^3}+
2\frac{F_3^{(s)}F_2^{(h)}F_3^{(1)}}{\langle N_s \rangle \langle N_h \rangle^2}+
\frac{F_3^{(s)}F_3^{(1)}F_2^{(1)}}{\langle N_h \rangle^3}+
2.5\frac{F_3^{(s)}F_3^{(h)}F_2^{(1)}}{\langle N_s \rangle^2 \langle N_h \rangle}+
\frac{F_3^{(h)}F_3^{(1)}F_2^{(1)}}{\langle N_s \rangle^2\langle N_h \rangle^3}\biggr]+ \nonumber \\
&& 45\biggl[
\frac{F_4^{(s)}F_2^{(1)2}}{\langle N_h \rangle^2}+
\frac{F_4^{(s)}F_2^{(h)2}}{\langle N_s \rangle^2}+
\frac{150}{45}\frac{F_4^{(s)}F_2^{(h)}F_2^{(1)}}{\langle N_s \rangle \langle N_h \rangle}+
\frac{F_4^{(h)}F_2^{(1)2}}{\langle N_s \rangle^3 \langle N_h \rangle^2}
\biggr]+ \nonumber \\
&& 15\biggl[\frac{F_3^{(s)}F_2^{(h)3}}{\langle N_s \rangle^3}+
\frac{F_3^{(s)}F_2^{(1)3}}{\langle N_h \rangle^3}+
\frac{F_3^{(h)}F_2^{(1)3}}{\langle N_s \rangle ^2 \langle N_h \rangle^3}+
18\frac{F_3^{(s)}F_2^{(h)}F_2^{(1)2}}{\langle N_s \rangle \langle N_h \rangle^2}+
15\frac{F_3^{(s)}F_2^{(h)2}F_2^{(1)}}{\langle N_s \rangle^2 \langle N_h \rangle}
\biggr]+ \nonumber \\
&& 180\biggl[
\frac{F_2^{(s)}F_3^{(h)}F_2^{(1)2}}{\langle N_s \rangle^2 \langle N_h \rangle^2}+
\frac{F_2^{(s)}F_2^{(h)}F_2^{(1)}F_3^{(1)}}{\langle N_s \rangle \langle N_h \rangle^3}
\biggr]+ \nonumber \\
&& 60\frac{F_2^{(s)}F_2^{(h)2}F_3^{(1)}}{\langle N_s \rangle^2 \langle N_h \rangle^2}+ 
135\frac{F_2^{(s)}F_2^{(h)2}F_2^{(1)2}}{\langle N_s \rangle^2 \langle N_h \rangle^2}+
45\frac{F_2^{(s)}F_2^{(h)}F_2^{(1)3}}{\langle N_s \rangle \langle N_h \rangle^3} \nonumber
\end{eqnarray}

\newpage




\begin{thebibliography}{999}

\bibitem{Haber:1984rc}
  H.~E.~Haber and G.~L.~Kane,
  Phys.\ Rept.\  {\bf 117} (1985) 75.

\bibitem{Strassler:2006im}
  M.~J.~Strassler and K.~M.~Zurek,
  Phys.\ Lett.\  B {\bf 651}, 374 (2007)
  [hep-ph/0604261].

\bibitem{Kang:2008ea}
 J.~Kang and M.~A.~Luty,
 JHEP {\bf 0911} (2009) 065
 [arXiv:0805.4642] [hep-ph]].


\bibitem{Alekhin:2015byh}
  S.~Alekhin {\it et al.},
  arXiv:1504.04855 [hep-ph].



\bibitem{Strassler:2008} 
  M.~J.~Strassler,
  arXiv:0806.2385 [hep-ph].




\bibitem{SanchisLozano:2008te}
  M.-A.~Sanchis-Lozano,
  Int.\ J.\ Mod.\ Phys.\ A {\bf 24} (2009) 4529
  [arXiv:0812.2397 [hep-ph]].


\bibitem{Sanchis-Lozano:2014goa}
  M.-A.~Sanchis-Lozano and E.~Sarkisyan-Grinbaum,
  arXiv:1409.5262 [hep-ph].


\bibitem{DeWolf:1995pc}
  E.~A.~De Wolf, I.~M.~Dremin and W.~Kittel,
  Phys.\ Rept.\  {\bf 270} (1996) 1
  [hep-ph/9508325].

\bibitem{Dremin:2000ep}
  I.~M.~Dremin and J.~W.~Gary,
  Phys.\ Rept.\  {\bf 349}, 301 (2001)
  [hep-ph/0004215].

\bibitem{Manjavidze:2001ni} 
  J.~Manjavidze and A.~Sissakian,
  Phys.\ Rept.\  {\bf 346}, 1 (2001)
  [hep-ph/0105245].

\bibitem{Kittel:book}
  For a comprehensive review on wide aspects of multiparticle 
production emphasising correlation studies, see: 
W.~Kittel,  E.~A.~De Wolf, 
  {\it Soft Multihadron Dynamics} (World Scientific, Singapore, 2005).


\bibitem{Dremin:2004ts}
I.~M.~Dremin and 
V.~A.~Nechitailo,
  Phys.\ Rev.\ D {\bf 70} (2004) 034005
  [hep-ph/0402286].


\bibitem{Dremin:2011sa}
  I.~M.~Dremin and V.~A.~Nechitailo,
  Phys.\ Rev.\ D {\bf 84} (2011) 034026
  [arXiv:1106.4959 [hep-ph]].




\bibitem{Dremin:1993sd}
  I.~M.~Dremin,
  Phys.\ Lett.\ B {\bf 313} (1993) 209.


\bibitem{Ugoccioni:1994zk}
  R.~Ugoccioni, A.~Giovannini and S.~Lupia,
  Phys.\ Lett.\ B {\bf 342} (1995) 387
  [hep-ph/9410340].


\bibitem{ppMultiplicity:Review} 
  J.~F.~Grosse-Oetringhaus and K.~Reygers,
  J.\ Phys.\ G {\bf 37}, 083001 (2010)
  [arXiv:0912.0023 [hep-ex]].



\bibitem{Giovannini:1985mz}
  A.~Giovannini and L.~Van Hove,
  Z.\ Phys.\  C {\bf 30} (1986) 391.




\bibitem{Fowler:1985fy}
  G.~N.~Fowler, E.~M.~Friedlander, R.~M.~Weiner and G.~Wilk,
 Phys.\ Rev.\ Lett.\  {\bf 56} (1986) 14.




\bibitem{Giovannini:1998zb}
  A.~Giovannini and R.~Ugoccioni,
  Phys.\ Rev.\ D {\bf 59} (1999) 094020
   [Phys.\ Rev.\ D {\bf 69} (2004) 059903]
  [hep-ph/9810446].



\bibitem{Giovannini:2003ft}
  A.~Giovannini and R.~Ugoccioni,
  Phys.\ Rev.\ D {\bf 68} (2003) 034009
  [hep-ph/0304128].


\bibitem{Zborovsky:2013tla} 
  I.~Zborovsk\'y,
  J.\ Phys.\ G {\bf 40} (2013) 055005
  [arXiv:1303.7388 [hep-ph]].

\bibitem{Ghosh:2013fsa} 
  P.~Ghosh and S.~Muhuri,
  Phys.\ Rev.\ D  {\bf 87} (2013) 094020
  [arXiv:1402.6820 [hep-ph]].

\bibitem{Adam:2015gka}
  J.~Adam {\it et al.} [ALICE Collaboration],
  arXiv:1509.07541 [nucl-ex].


\bibitem{Kaidalov:1982xe}
  A.~B.~Kaidalov and K.~A.~Ter-Martirosyan,
  Phys.\ Lett.\ B {\bf 117} (1982) 247.


\bibitem{Matinyan:1998ja}
  S.~G.~Matinyan and W.~D.~Walker,
  Phys.\ Rev.\ D {\bf 59} (1999) 034022
  [hep-ph/9801219].

\bibitem{Beggio:2013vfa}
  P.~C.~Beggio and E.~G.~S.~Luna,
  Nucl.\ Phys.\ A {\bf 929} (2014) 230
  [arXiv:1308.6192 [hep-ph]].




\bibitem{Dremin:1994wm}
I.~M.~Dremin,
Phys.\ Lett.\ B {\bf 341} (1994) 95, [Phys.\ Lett.\ B {\bf 348} (1995) 711 (E)]
[hep-ph/9408300].


\bibitem{Buican:2003jx}
  M.~A.~Buican, C.~Forster and W.~Ochs,
  Eur.\ Phys.\ J.\ C {\bf 31} (2003) 57
  [hep-ph/0307234].



\bibitem{Khoze:1996dn}
  V.~A.~Khoze and W.~Ochs,
  Int.\ J.\ Mod.\ Phys.\ A {\bf 12} (1997) 2949
  [hep-ph/9701421].


\bibitem{Abe:1996vs} 
  K.~Abe {\it et al.} [SLD Collaboration],
  Phys.\ Lett.\ B {\bf 371}, 149 (1996)
  [hep-ex/9601010].

\bibitem{Achard:2001ut} 
  P.~Achard {\it et al.} [L3 Collaboration],
  Phys.\ Lett.\ B {\bf 577}, 109 (2014)
  [hep-ex/0110072].




\bibitem{Dremin:1997hk} 
  I.~M.~Dremin, V.~A.~Nechitailo, M.~Biyajima and N.~Suzuki,
  Phys.\ Lett.\ B {\bf 403}, 149 (1997)
  [hep-ph/9704318].



\bibitem{Sarkisian:2000ux} 
  E.~K.~G.~Sarkisyan,
  Phys.\ Lett.\ B {\bf 477}, 1 (2000)
  [hep-ph/0001262].

\bibitem{Abbiendi:1999ki} 
  G.~Abbiendi {\it et al.} [OPAL Collaboration],
  Eur.\ Phys.\ J.\ C {\bf 11}, 239 (1999)
  [hep-ex/9902021].


\bibitem{Abbiendi:2001bu} 
  G.~Abbiendi {\it et al.} [OPAL Collaboration],
  Phys.\ Lett.\ B {\bf 523}, 35 (2001)
  [hep-ex/0110051].


\bibitem{E735}
  E735 Collaboration, F. Turkot et al., Nucl.\ Phys\ A {\bf 525} (1991) 165

\bibitem{prolog}  

W.F. Clocksin and C.S. Mellish, {\em Programming in Prolog}, 5th edition. 2003. 
Springer-Verlag. ISBN 3-540-00678-8  

\bibitem{prolog:2015}
M.-A. Sanchis-Lozano, E. K. Sarkisyan-Grinbaum, S. Moreno-Picot,  paper in 
preparation



\end{thebibliography}







\end{document}